\documentclass[floatfix,aps,pra,twocolumn,showpacs,superscriptaddress,10pt]{revtex4-2}
\usepackage[utf8]{inputenc}
\usepackage[T1]{fontenc}
\usepackage{color}
\usepackage{float}
\usepackage{mathrsfs}
\usepackage{amstext}
\usepackage{amsmath}
\usepackage{graphicx}
\usepackage{epstopdf}

\usepackage{lipsum}

\usepackage{babel}
\begin{document}
\preprint{APS/123-QED}
\title{Highly sensitive controllability of optical bistability in three-level atomic systems}
\author{M. H. Oliveira}
\email{murilo.oliveira@df.ufscar.br }

\affiliation{Departamento de Física, Universidade Federal de São Carlos, P.O. Box 676, 13565-905, São Carlos, São Paulo, Brazil. }
\author{H. S. Borges}
\email{halyneborges@iftm.edu.br }
\affiliation{Instituto Federal do Triângulo Mineiro, Campus Patrocínio, 38740-000, Patrocínio, Minas Gerais, Brazil. }

\author{J. A. Souza}
\email{jasouza@ufscar.br }
\affiliation{Departamento de Física, Química e Matemática, Universidade Federal de São Carlos,18052-780, Sorocaba, São Paulo, Brazil.}

\author{C. J. Villas-Boas}
\email{celsovb@df.ufscar.br }
\affiliation{Departamento de Física, Universidade Federal de São Carlos, P.O. Box 676, 13565-905, São Carlos, São Paulo, Brazil. }

\date{\today}
\begin{abstract}
We theoretically investigate the optical bistability phenomenon in an ensemble of $N$ non-interacting three-level atoms trapped inside an optical cavity. The atoms are in a $\Lambda$-level configuration, where one atomic transition is coupled by a cavity mode, while the other one is coupled by a classical field. In addition, we consider a pumping field driving the cavity mode. With this system, we are able to observe new kinds of hysteresis, while scanning either the frequency of the pumping field or the Rabi frequency (intensity) of the control field. We show that they can be highly controllable via external parameters of the system, achieving very narrow widths, thus being very useful for building new devices, such as small fluctuation detectors in either frequency or intensity of laser fields.

\end{abstract}

\maketitle

\section{\label{sec:1}Introduction}

The optical bistability (OB) phenomenon observed in atomic systems occurs when there are two different values of transmission for a single value of driving field intensity input. This happens since the system can present two stable states, due to the nonlinear character of the interaction between the radiation field and the absorptive medium. OB can be observed in an atomic ensemble confined in a Fabry-Perot resonator, which is one of the most fundamental physical systems where the effects associated to the light-matter interaction have been extensively studied in the last decades. The OB phenomenon was predicted by Szöke \textit{et al.} \citep{Szoke69}, where they proposed the implementation of a bistable device which could be useful as a switching element in optical communications. Szöke \textit{et al.} studied the nonlinear behavior of an irradiated gas, acting as a saturable absorber confined inside a Fabry-Perot resonator. A few years after, in 1976, Gibbs \textit{et al.} accomplished the first experimental observation of the OB phenomenon \citep{GibbsMcCall76}. Since then, almost five decades later, the optical bistability is still an intensely investigated subject in different physical systems \citep{Bloch17,Rodriguez20,lsun2018,lsun2019,lsun2021,parmee2021,jeannin2021}, primarily due to the fact it is closely related to other interesting optical phenomena such as the implementation of laser amplifiers \citep{Wang17}, sensing elements \citep{Rodriguez20}, all-optical switches \citep{Yadipoura18} and optical quantum memories \citep{mohammadi2022}.

In systems of two or three-level atoms trapped inside an optical cavity, the OB phenomenon is characterized by the appearance of a hysteresis loop in the transmission curve as a function of the input intensity laser. For two-level atoms, it is possible to obtain an analytical solution for the steady state of the system, which can exhibit a nonlinear character. Such nonlinear behavior can be only observed for $C>4$, being $C$ the Cooperativity parameter. For three-level atoms, however, this threshold is not well established, since there is no steady state analytical solution for this case. 

Multilevel systems, where transitions are coupled by different laser fields, lead to a vast number of controllable optical phenomena and applications such as the absorptive bistability \citep{Walls1980,Walls1981,Lawandy1984,Wang2002,Joshi2003a,Brown2003,Joshi2003,Li2008,Joshi2010,Wu2010,Wang2012,Vafafard2013,Wang2013}. The presence of a control field, for instance, can induce quantum interference and coherent effects that enhance the aspects of the system's nonlinear optical response. In $\Lambda$-type three-level atoms, under the conditions of the electromagnetically induced transparency (EIT) regime~\cite{ImamogluRev2005}, the OB effect can be directly controlled, since the hysteresis curve can be properly manipulated by tuning external physical parameters, such as the Rabi frequency of control field \citep{Lawandy1984,Wang2002,Li2008,Wu2010,Joshi2010,Wang2012,Vafafard2013}. Although Walls and Zoller \cite{Walls1980} investigated the characteristic hysteresis of $\Lambda$-type three-level atomic system as a function of the probe detuning, they did it for a very unique system, where both atomic transitions were coupled to the same cavity mode with symmetric detunings, what leads to some key simplifications in the system's Bloch equations. To the best of our knowledge, no previous work considering a general system of three-level atoms have analyzed the hysteresis curve as a function of the Rabi frequency of the control field or even as a function of the probe frequency detuning, as formerly investigated in two-level systems \citep{Ritter2009,Shirai2018}. 

In this paper, we theoretically investigate the optical bistability phenomenon in a sample of non-interacting three-level atoms confined inside an optical cavity. Taking advantage of the high controllability of this system's bistable behavior, both as a function of the cavity-probe detuning and the intensity control laser, we propose the implementation of devices capable of measuring very small fluctuations of intensity and frequency of laser fields. Based on our numerical results, we also studied the minimum Cooperativity in which the three-level atoms display a bistable character, which is the key parameter for the OB and, consequently, for the development of optical sensors. Unlike for two-level systems, here we show that three-level $\Lambda$-systems can present OB even for $C<4$.

\section{\label{sec:2}Physical System and Model }

\begin{figure}
\includegraphics[width=1\columnwidth]{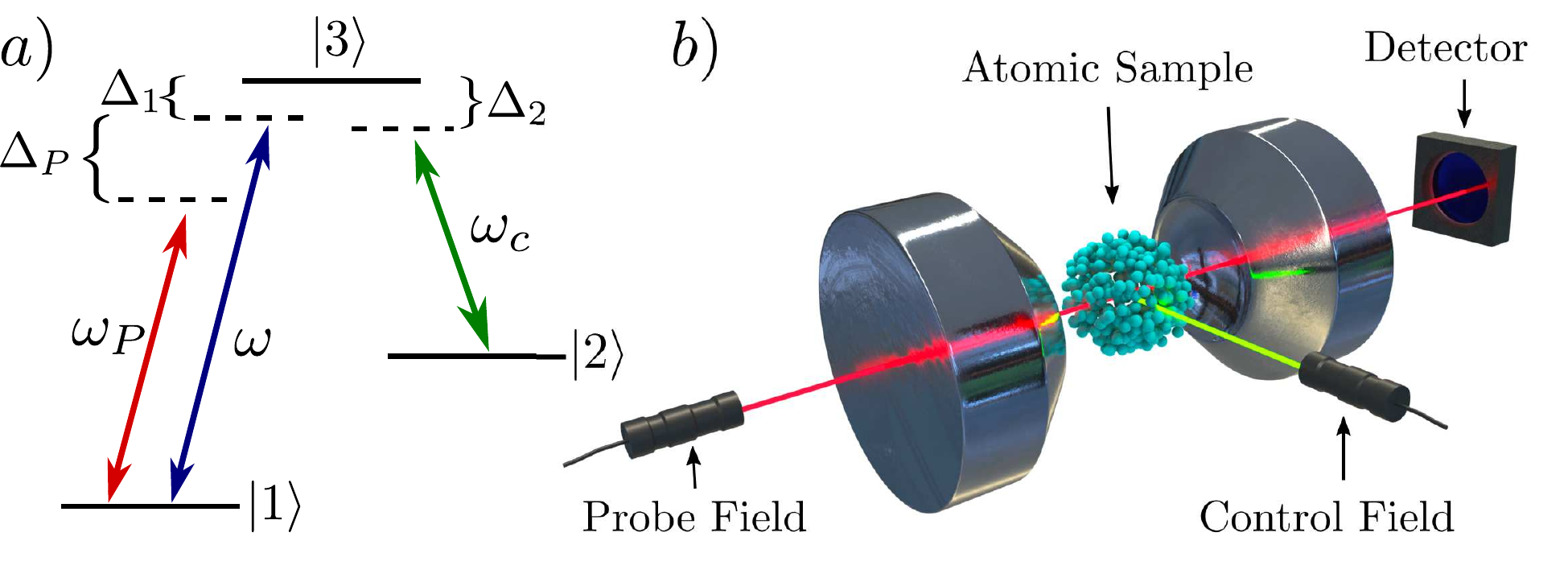}
\caption{(Color online) a) Energy levels diagram of three-level atoms in a $\Lambda$-configuration trapped inside an optical cavity. The atom-cavity system is pumped by a probe laser with frequency $\omega_{P}$ and strength $\varepsilon$. The transition $|2\rangle\rightarrow|3\rangle$ is coupled by a control classical field with frequency $\omega_{C}$ and Rabi frequency $2\Omega_{C}$. The cavity mode, of frequency $\omega$, couples the transition $|1\rangle\rightarrow|3\rangle$ with coupling strength $g$. b) Artist's view of the experimental apparatus used for observation the cavity-EIT phenomenon and the controllable optical bistability, subject to the incidence of the control and probe fields, with a light detector properly positioned in order to measure the cavity transmission. 
}
\label{fig:1}
\end{figure}

Consider $N$ non-interacting atoms (very diluted gas such that light-mediated interactions can be neglected~\cite{Oliveira2021}) confined into a two-sided optical cavity, with one of its sides pumped by a coherent probe laser. Each atom in the sample is described as a three-level system in a $\Lambda$-level configuration, being the excited state $|3\rangle$ coupled to the ground state $|1\rangle$ via the intracavity mode (frequency $\omega$), with coupling strength $g$, while the other ground state $|2\rangle$ is coupled to the excited state $|3\rangle$ by a classical control field (frequency $\omega_{c}$) with Rabi frequency $2\Omega_{c}$. We also consider a pumping field acting onto the intracavity mode with strength $\varepsilon$ and frequency $\omega_{P}$. The energy levels of each atom, as well as the artist's view of the experimental setup is shown in Fig. \ref{fig:1}.

Under the dipole interaction, rotating-wave and the Born-Markov approximations, the dynamics of the density matrix $\rho$ of the system investigated here can be described by the following master equation (at $T=0$ K and with $\hbar=1$)
\begin{eqnarray}
\frac{d\rho}{dt} & = & -i[H,\rho]+\mathcal{L}_{at}(\rho)+\mathcal{L}_{cav}(\rho),\label{eq:2}
\end{eqnarray}
where the Hamiltonian that describes the system, in a time-independent rotating frame, is \cite{CelsoEITprl}
\begin{eqnarray}
H & = & \Delta_{1}S_{33}+(\Delta_{1}-\Delta_{2})S_{22}+\Delta_{P}S_{11}-\Delta_{P}a^{\dagger}a\nonumber \\
 &  & +(gaS_{31}+\Omega_{C}S_{32}+\varepsilon a)+H.c.,\label{eq:1}
\end{eqnarray}
where $S_{ij}=\sum_{k=1}^{N}|i\rangle^{(k)}\langle j|^{(k)}=\sum_{k=1}^{N}\sigma_{ij}^{(k)}$ is the collective raising and lowering atomic operator. The intracavity mode is represented by the photon annihilation $a$ and creation $a^{\dagger}$ operators, and the detunings are given by $\Delta_{1}=\omega_{31}-\omega$, $\Delta_{2}=\omega_{32}-\omega_{C}$, being $\omega_{3l}$ the atomic transition frequency between the excited $\vert 3 \rangle$ and ground state $\vert l\rangle$, with $l=1,2$. $\Delta_{P}=\omega_{P}-\omega$ is the detuning between the probe field and the single mode cavity frequencies. Finally, $H.c.$ stands for the Hermitian conjugate. For simplicity we consider the case of a perfect cavity and control field resonance $\Delta_{1}=\Delta_{2}=0$.
\begin{align*}
\mathcal{L}_{at}(\rho) & =\sum_{k=1}^{N}\sum_{l=1,2}\Gamma_{3l}\left(2\sigma_{l3}^{\left(k\right)}\rho\sigma_{3l}^{\left(k\right)}-\sigma_{33}^{\left(k\right)}\rho-\rho\sigma_{33}^{\left(k\right)}\right)\\
 & +\sum_{k=1}^{N}\sum_{j=2,3}\gamma_{j}\left(2\sigma_{jj}^{\left(k\right)}\rho\sigma_{jj}^{\left(k\right)}-\sigma_{jj}^{\left(k\right)}\rho-\rho\sigma_{jj}^{\left(k\right)}\right)
\end{align*}
is the Liouville operator that describes the dissipation of each atom, with $\Gamma_{3l}$ and $\gamma_{j}$ representing the polarization decay rate of the excited level $|3\rangle$ to the ground state $|l\rangle$ and the pure dephasing rate of the level $|j\rangle$, respectively. The Liouville operator
\[
\mathcal{L}_{cav}(\rho)=\kappa\left(2a\rho a^{\dagger}-a^{\dagger}a\rho-\rho a^{\dagger}a\right)
\]
describes the cavity field dissipation, with $\kappa$ being the total decay rate of the amplitude of the cavity field.

The number of atoms trapped inside the cavity is a very important physical parameter to the study of the optical bistability phenomenon, since the bistable behavior of the system is a nonlinear effect which usually manifests itself for a large atomic Cooperativity, defined as $C\equiv Ng^{2}/2\kappa\Gamma_{3}$, being $\Gamma_{3}=\Gamma_{31}+\Gamma_{32}$ the total decay rate of the excited state. However, for large number of atoms, $N\gg 1$, the dimension of the relevant Hilbert space increases drastically, what makes the solution of a complete quantum model intractable, even with a supercomputer. In order to circumvent this problem, we work in a weak coupling regime for the atom-cavity system, i.e., $g<(\kappa,\Gamma_{3l})$. In general, in our simulations we used $N\sim10^{^{3}}-10^{^{4}}$ atoms and $g\sim10^{^{-1}}\kappa$. The main advantage of restricting ourselves to this regime is that the dissipative dynamics of the system can be solved using a semi-classical (SC) approximation, which allows us to solve the equations for a large number of atoms. In the regime we are interested here, the atoms act as a classical nonlinear intracavity medium which is basically the semi-classical understanding for the absorptive optical bistability. 

To investigate the OB phenomenon in the three-level system, we must calculate the average number of photons inside the cavity $\langle n \rangle = \langle a^{\dagger} a \rangle $. Using the property $\langle\dot{O}\rangle = Tr (\dot{\rho}O)$, with $O$ being any atomic or cavity field operator and $Tr$ the trace of the resulting matrix, it is possible to obtain its temporal evolution through the master equation (\ref{eq:2}). Following this procedure, we end up an infinite linear system of coupled differential equations. For $N\gg 1$ and assuming the weak coupling regime limit, we can neglect the atom-field correlations, i.e. $\langle aS_{ij}\rangle\approx\alpha\langle S_{ij}\rangle$, which is a good approximation in the classical limit, where we replaced $\langle a\rangle\rightarrow\alpha$ ($\langle a^{\dagger}\rangle\rightarrow\alpha^{*}$), where $\alpha$ is a time dependent amplitude of the cavity field. Applying this semi-classical approximation to the infinite system of linear equations previously obtained, turns it into a finite nonlinear system of coupled differential equations, which reads as
\begin{subequations}
\begin{eqnarray}
\dot{\alpha} & = & i\left(\Delta_{P}+i\kappa\right)\alpha-i\varepsilon^{\ast}-ig^{\ast}\left\langle 
\label{se1}
S_{13}\right\rangle ,\\
\langle\dot{S_{13}}\rangle & = & i\left\{ \left(\Delta_{P}-\Delta_{1}\right)+i\left(\Gamma_{31}+\Gamma_{32}+\gamma_{3}\right)\right\} \left\langle S_{13}\right\rangle \nonumber \\
 &  & -i\Omega_{c}\left\langle S_{12}\right\rangle +ig\alpha\left(\left\langle S_{33}\right\rangle -\left\langle S_{11}\right\rangle \right),\\
\langle\dot{S_{12}}\rangle & = & i(\Delta_{P}+\Delta_{2}-\Delta_{1}+i\gamma_{2})\left\langle S_{12}\right\rangle \nonumber \\
 &  & -i\Omega_{c}^{\ast}\left\langle S_{13}\right\rangle +ig\alpha\left\langle S_{32}\right\rangle ,\\
\langle\dot{S_{23}}\rangle & = & i\left\{ -\Delta_{2}+i\left(\Gamma_{31}+\Gamma_{32}+\gamma_{2}+\gamma_{3}\right)\right\} \left\langle S_{23}\right\rangle \nonumber \\
 &  & -ig\alpha\left\langle S_{21}\right\rangle +i\Omega_{c}\left(\left\langle S_{33}\right\rangle -\left\langle S_{22}\right\rangle \right),\\
\langle\dot{S_{11}}\rangle & = & -ig^{\ast}\alpha^{\ast}\left\langle S_{13}\right\rangle +ig\alpha\left\langle S_{31}\right\rangle +2\Gamma_{31}\left\langle S_{33}\right\rangle ,\\
\langle\dot{S_{22}}\rangle & = & -i\Omega_{c}^{\ast}\left\langle S_{23}\right\rangle +i\Omega_{c}\left\langle S_{32}\right\rangle +2\Gamma_{32}\left\langle S_{33}\right\rangle ,\\
\langle\dot{S_{33}}\rangle & = & -\langle\dot{S_{11}}\rangle-\langle\dot{S_{22}}\rangle.
\label{se9}
\end{eqnarray}
\end{subequations}

As mentioned before, the majority of studies published in the last decades about OB consider two-level systems, where it is possible to obtain a nonlinear analytical expression for the state equation of the system in the stationary regime, from which, we are able to investigate the bistable character of the transmitted field as a function of the input field or the detuning between the atomic transition and the mode cavity \citep{Bonifacio1976,Joshi2010}. When considering the three-level system, however, the resulting system of equations is no longer analytically solvable. Thus, in this work, the system of equations was solved using two numerical methods: the first one consists in integrating the differential equations for a sufficiently long time so that the system could reach the steady state; the second one consists in finding the stationary solution of the system numerically, i.e., setting the time derivative of the operators' average equal to zero in the Maxwell-Bloch equations. The results presented here were obtained implementing both methods in order to ensure consistency.

\section{\label{sec:3} Bistable behavior in the three-level atomic system}

\begin{figure}
\includegraphics[width=1\columnwidth]{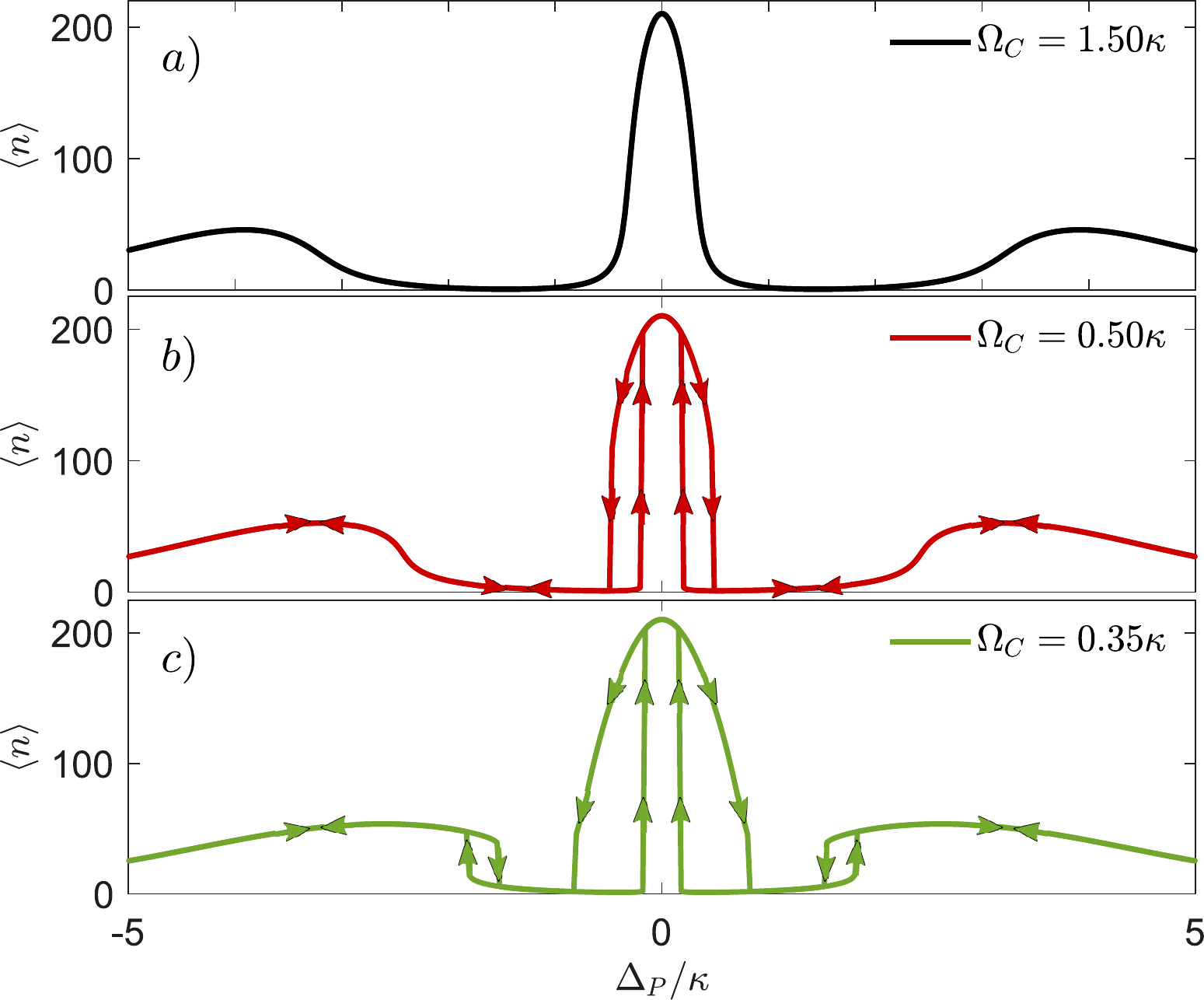}

\caption{(Color online) Average number of intracavity photons $\langle n \rangle$ versus the normalized detuning $\Delta_P/\kappa$ for different values of the control field Rabi frequency, showing three possible regimes one could achieve: in a) there is no bistability at all ($\Omega_C = 1.5\kappa$), in b) the bistability is solely related to the EIT dark state ($\Omega_C = 0.5\kappa$), in c) we can have both situations, i.e., bistability due to normal modes (two-level signature) and due to EIT dark states ($\Omega_C = 0.35\kappa$). When solving time dynamics, we integrate the equations of motion for each value of $\Delta_P$ by taking as the initial conditon the steady state obtained for the immediately preceding $\Delta_P$ value. Thus, the arrows in figures b) and c) indicate the direction in which we vary the $\Delta_p$ (increasing or decreasing). Parameters used: $C=8$, $N=10^3$, $\Gamma_{31}=\Gamma_{32}=0.5\kappa$ and $\varepsilon=14.5\kappa$.}
\label{fig:1_1}
\end{figure}

When multilevel atoms are confined into an optical cavity they present a great advantage compared to two-level systems, since the optical response can be properly manipulated by additional external parameters, such as the frequency and the intensity of a second laser field that couples the additional atomic transition \citep{Walls1980,Walls1981,Lawandy1984,Wang2002,Joshi2003a,Brown2003,Joshi2003,Li2008,Joshi2010,Wu2010,Wang2012,Vafafard2013,Wang2013}. In the three-level $\Lambda$-type system, nonlinear optical processes are substantially affected by the presence of the control field. In these atoms, under the influence of a probe and control field, it is possible to observe nonlinear mechanisms such as coherent population trapping (CPT), EIT and Kerr-nonlinear index of refraction \citep{ImamogluRev2005}. Fig. \ref{fig:1_1} shows the average number of photons $\langle n \rangle $ in the cavity mode as a function of the probe field detuning $\Delta_P/\kappa$ for different values of the control field Rabi frequency $\Omega_c$. It is important to notice that there are two distinct bistable regions: one around the normal mode peaks, which is related to the usual bistability in two-level atoms, and another around the EIT peak, which appears only in three-level systems. 

\begin{figure}
\includegraphics[width=1\columnwidth]{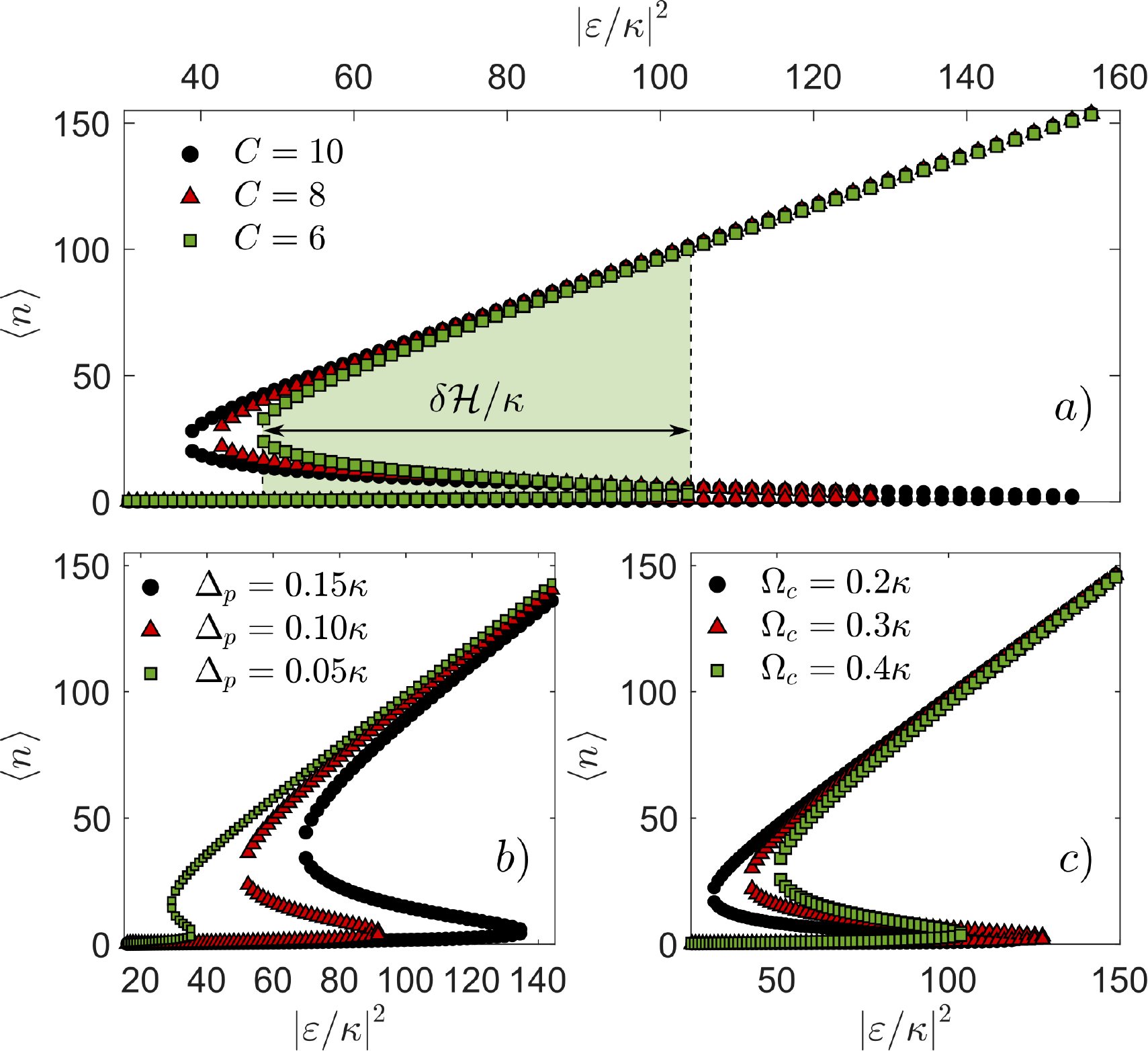}
\caption{(Color online) Average number of photons $\langle n\rangle=\left|\alpha\right|^{2}$ inside the cavity as a function of $|\varepsilon/\kappa|^{2}$ for $N=10^{3}$ atoms, which are considered to be $\Lambda$-type three-level systems under the incidence of probe and control fields for different parameter configurations: a) $\Delta_{P}=0.1\kappa$, $\Omega_{C}=0.3\kappa$, and different values for the Cooperativity $C$. The light green filling highlights the effective area of the hysteresis obtained for $C=6$, and $\delta\mathcal{H}/\kappa$ indicates its characteristic width, which can be controlled by adjusting the parameters; b) $C=5$, $\Omega_{C}=0.3\kappa$, and different values for the cavity detuning $\Delta_{P}$; c) $C=8$, $\Delta_{P}=0.1\kappa$, and different values for the Rabi frequency of the control field $\Omega_{C}$. The other physical parameters are: $\Gamma_{31}=\Gamma_{32}=0.5\kappa$.}

\label{fig:2}
\end{figure}

 In order to analyze how the OB can be controlled and then discuss how this system can be used as detectors of small fluctuations in frequency and intensity of laser fields, it is crucial to evaluate under which parameter regime the system presents a bistable character and investigate how the shape of the OB hysteresis curve is modified as a function of the external parameters. To this end, first we analyze the characteristic hysteresis for $\Lambda$-type three-level atoms as a function of the probe field input intensity. This kind of hysteresis was already investigated experimentally \citep{Wang2002,Brown2003,Joshi2003a,Joshi2010}. Fig. \ref{fig:2} shows the average number of photons inside the cavity as a function of $|\varepsilon/\kappa|^{2}$, for different values of the Cooperativity parameter $C$, $\Delta_{P}$, and $\Omega_{C}$. As the dephasing rates are usually much smaller than the polarization decay rates, from now on we fixed $\gamma_2 = \gamma_3 = 0$.

Just as in the case of two-level systems, the larger the Cooperativity value $C$, the wider the hysteresis, as seen in Fig. \ref{fig:2}(a). Here, this parameter is changed by varying the coupling strength $g$, and keeping all other parameters fixed. Fig. \ref{fig:2}(b) and (c) also shows that the hysteresis becomes wider as the cavity-probe detuning $\Delta_{P}$ is increased and the control field Rabi frequency $\Omega_{C}$ is decreased. For a (small) null $\Delta_{P}$ the system is in a (almost) perfect EIT configuration, thus being completely transparent for the probe field, resulting in the absence of hysteresis. The hysteresis also tends to disappear for large $\Omega_{C}$ since the stronger the control field Rabi frequency, the more transparent the system becomes. All curves in Fig. \ref{fig:2} show an average number of photons $\gg1$. This behavior is also present in two-level atomic systems, since the hysteresis appears only in the nonlinear regime, reached when the system saturates. In both cases, i.e., with two or three-level atoms, this kind of hysteresis is directly related to normal modes of the atom-field coupling of the system. However, it is also possible to achieve the bistable regime in the non resonant driving regime, as it was showed for two-level atoms by S. Ritter \textit{et al.} \citep{Ritter2009} and by T. Shirai \textit{et al.} \citep{Shirai2018}, and the resulting hysteresis curve appears in the transmission spectrum of the cavity as a function of the pump-cavity detuning even for a very low intracavity photon number. As in the two-level systems, the transmitted field from the $\Lambda$-type three-level atom-cavity system can also exhibit a bistable behavior as a function of parameters other than the input field strength, which, to the best of our knowledge, was not investigated so far. 

\begin{figure}
\includegraphics[width=1\columnwidth]{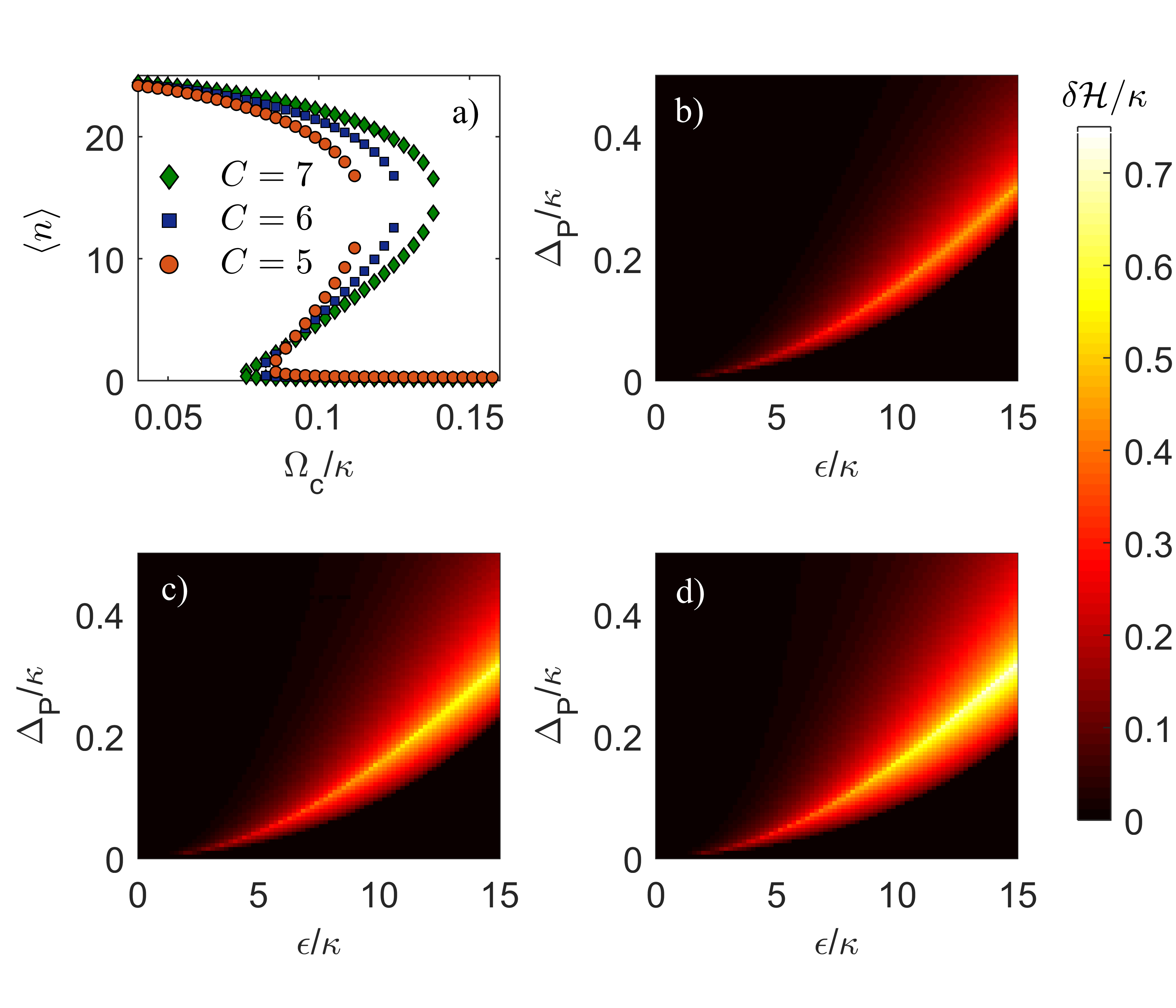}

\caption{(Color online) (a) Average number of intracavity photons $\langle n\rangle$ as a function of Rabi frequency of the control field $\Omega_{C}$ for different values of the Cooperativity parameter $C$, considering $\Delta_{P}=0.1\kappa$ and $\varepsilon=5\kappa$. (b), (c) and (d) are color maps that provide the measurement of the hysteresis width $\delta\mathcal{H}/\kappa$ as a function of each ordered pair of the parameters $\varepsilon$ and $\Delta_{P},$ and for Cooperativities $C=5$, $C=6$, and $C=7$, respectively. The other parameters considered here are: $N=10^3$ and $\Gamma_{31}=\Gamma_{32}=0.5\kappa$.}

\label{fig:3}
\end{figure}

From now on, let's investigate the new kinds of hysteresis in $\Lambda$-type three-level atom-cavity system, i.e., as a function of the detuning $\Delta_{P}$ and the Rabi frequency $\Omega_{C}$. The profile of the hysteresis curve as a function of the $\Omega_{C}$ is showed in the Fig.~\ref{fig:3}(a) for $\Delta_{p}=0.1\kappa$ and $\varepsilon=5\kappa$, assuming different values of Cooperativity parameter ($C=5,6$ and $7$). The panels \ref{fig:3}(b), (c) and (d) show the hysteresis width $\delta\mathcal{H}/\kappa$ (see Fig. \ref{fig:2}(a)) for different values of $C$, represented in a color map for each pair of the $\varepsilon$ and $\Delta_{P}$ parameters. These results allow us to characterize the set of parameters in which the system is bistable. The arising of the hysteresis can be associated either to the normal mode splitting, for strong cooperativities, or to the dark state of the system, which in turn, is dependent on the probe-cavity detuning. For a fixed value of $\varepsilon$, when $\Delta_{P}=0$, the system is completely transparent to the probe field and then, there is no hysteresis. For small values of $\Delta_{P}$, there will be a cavity-probe frequency range, for a fixed value of $\varepsilon$, where the bistable behavior arises. As for higher values of $\Delta_{P}$ the saturation of the system is not reached. These results also show that the maximum value of the hysteresis width increases for greater values of Cooperativity.

\begin{figure}
\includegraphics[width=1\columnwidth]{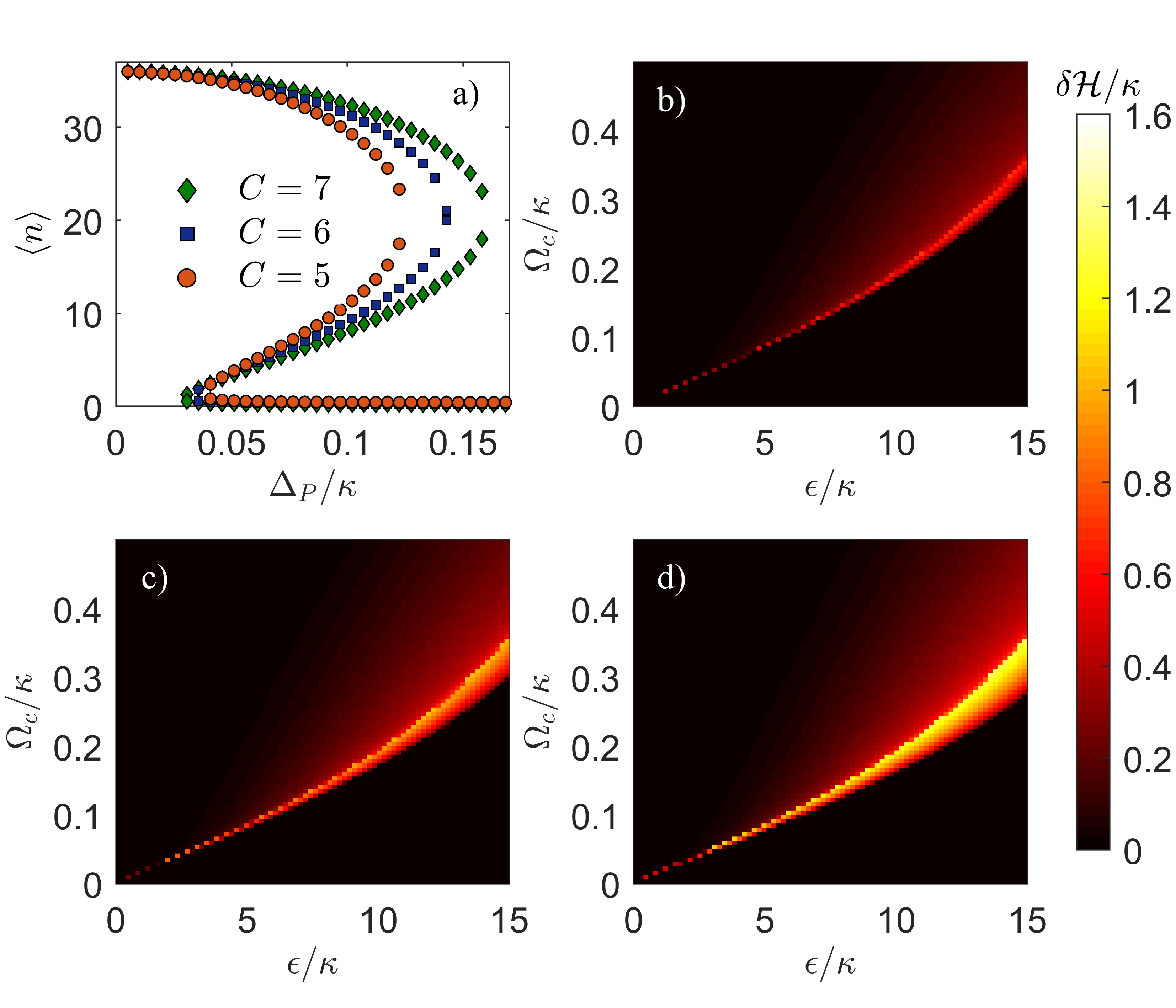}

\caption{(Color online) (a) Average number of intracavity photons $\langle n\rangle$ as a function of probe-cavity detuning $\Delta_{P}$ for different values of the Cooperativity parameter $C$, considering $\Omega_{C}=0.15\kappa$ and $\varepsilon=6\kappa$. (b), (c), and (d) are color maps that provide the measurement of the hysteresis width $\delta\mathcal{H}/\kappa$ as a function of each ordered pair of parameters $\varepsilon$ and $\Delta_{P}$ for Cooperativities $C=5$, $C=6$ and $C=7$, respectively. The other parameters considered here are the same considered in the Fig. \ref{fig:3}.}

\label{fig:4}
\end{figure}

A similar analysis is done in the Fig. \ref{fig:4}(a), where the OB hysteresis curve is plotted as a function of probe-cavity detuning $\Delta_{P}$. Panels \ref{fig:4}(b), (c), and (d) show the regions associated to a range of values of the input probe and control field intensity for which the system exhibits a bistable character. Analogously to Fig.~\ref{fig:3}, this analysis was carried out for $C=5,6$ and $7$, and the color maps show a very similar behavior for the width of the hysteresis as a function of the external parameters $\Omega_C$ and $\varepsilon$. However, in this case, the hysteresis curves are much more sensitive to the variations of these parameters, as we can see in the color maps, where the bistable regions are narrower and yet show a much larger variation in width. That means that with a very fine tuning of parameters, hysteresis with widths of different orders of magnitude can be obtained.

The simulations show that through the appropriate control and manipulation of external parameters of the system, as $\Omega_{C}$, $\varepsilon$ and $\Delta_{P}$, it is possible to control the OB hysteresis curve with considerable precision, being able to increase, decrease or even move the hysteresis curve associated to the atomic optical bistability of the system. Thus, one can properly prepare the system so that it presents a very specific bistable character, which in turn can be chosen according to the intended application. 

In this context, it is still important to analyze the critical Cooperativity in which the OB can not be observed, since the minimum value of this parameter establishes the threshold region where the fluctuation detectors can be implemented. By considering a single cavity mode coupling both atomic transitions, i.e., without a control field, it is possible to derive analytical expressions for the average number of photons and thus, the smallest Cooperativity required to observe bistability, which is $C\geq 1$ \cite{Walls1980}. This value is smaller than the one which appears in two-level systems ($C>4$), due to the high nonlienarity of this system: by coupling the same cavity mode with two different atomic transitions, the nonlinear effects become stronger. However, as far as we know, it is not possible to find an analytical solution for the steady state for the system in the general case (with a control field coupling the transition $|2\rangle\leftrightarrow|3\rangle$), as the one described here. Thus, the characterization of the bistable regime for our system can only be done through numerical simulations. By doing so, the minimum value of the Cooperativity in which the hysteresis curve appears was investigated. This analysis is accomplished by numerically calculating the hysteresis width as a function of the other parameters ($\varepsilon$, $\Delta_{P}$ or $\Omega_{C}$), considering different values of Cooperativity. 

\begin{figure}
\includegraphics[width=1\columnwidth]{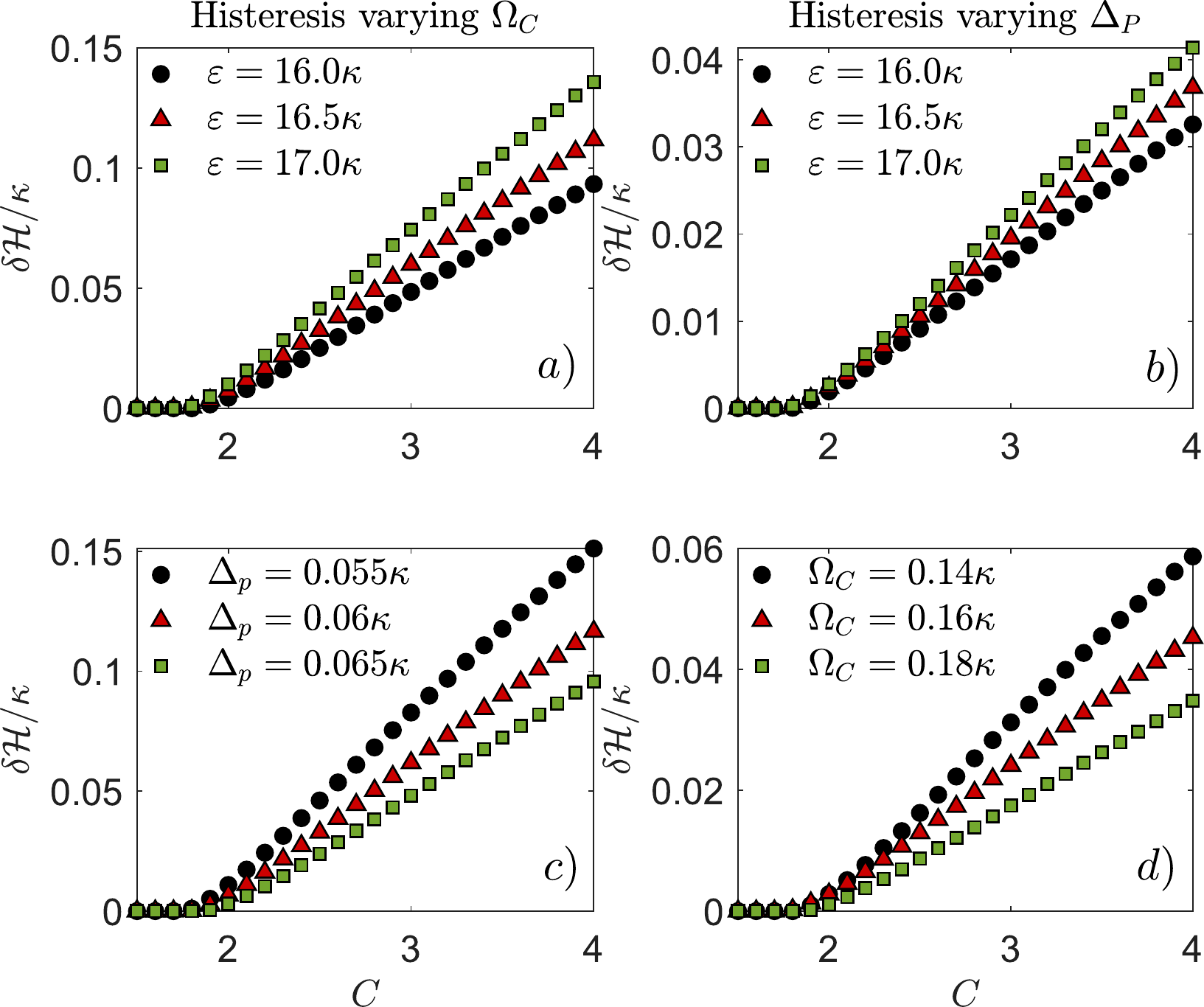}

\caption{(Color online) Bistability threshold for three-level systems. In panels a) and c), the normalized hysteresis width ($\delta\mathcal{H}/\kappa$) when varying the control field Rabi frequency $\Omega_C$ is measured for different values of Cooperativity and then for different probe field strengths and detunings, respectively. In panels b) and d), the same measurement is taken for the hysteresis observed when varying the probe field detuning, for different values of probe field strengths and control field Rabi frequencies, respectively. The number of atoms was fixed as $N=10^4$ in all the simulations. In a) $\Delta_P=0.05\kappa$, in b) $\Omega_C=0.15\kappa$ and in both c) and d) $\varepsilon=18\kappa$.}

\label{fig:5}
\end{figure}

In the Fig. \ref{fig:5} we show the normalized hysteresis width ($\delta\mathcal{H}/\kappa$) curve for different Cooperativity values. In (a) and (c) the hysteresis curves are obtained as a function of $\Omega_C$, and in (b) and (d) as a function of $\Delta_P$. All the hysteresis were obtained using $N=10^4$. For each panel, one of the two remaining free external parameters was fixed while the same curve was calculated for different values of the second one, e.g. in panel (a), $\Delta_{P}$ was fixed and three curves for different $\varepsilon$ were obtained.  Given the range of parameters investigated here, our results show hysteresis for cooperativities as low as $C=1.75$, which is below the threshold for bistability in two-level systems.  However, this result perhaps does not reflect the minimum value of $C$ in order to observe bistability in such systems, since through a careful manipulation of the system's parameters it could be possible to lower this minimum value even further. But, speaking of practical implementations, these hysteresis will become narrower while paying the price of losing contrast between the maximum and minimum transparency.  The global smallest value of Cooperativity which allows for bistablility in our system is still an open question, which could be solved by finding an analytical solution for the system of equations (\ref{se1}) to (\ref{se9}).

\section{Measurements of small fluctuations using atom-cavity system detectors}

In order to implement detector devices of small frequency and intensity fluctuations, using as key ingredient the controllability of the bistable behavior of the $\Lambda$-type three-level system, it is important to prepare the system according to the desired detection precision. Thus, depending on which parameter one wish to detect the fluctuation, the knowledge of the bistable behavior of the system acquired through the analysis made in the previous section is essential for the development of the detectors.

\begin{figure}
\includegraphics[width=1\columnwidth]{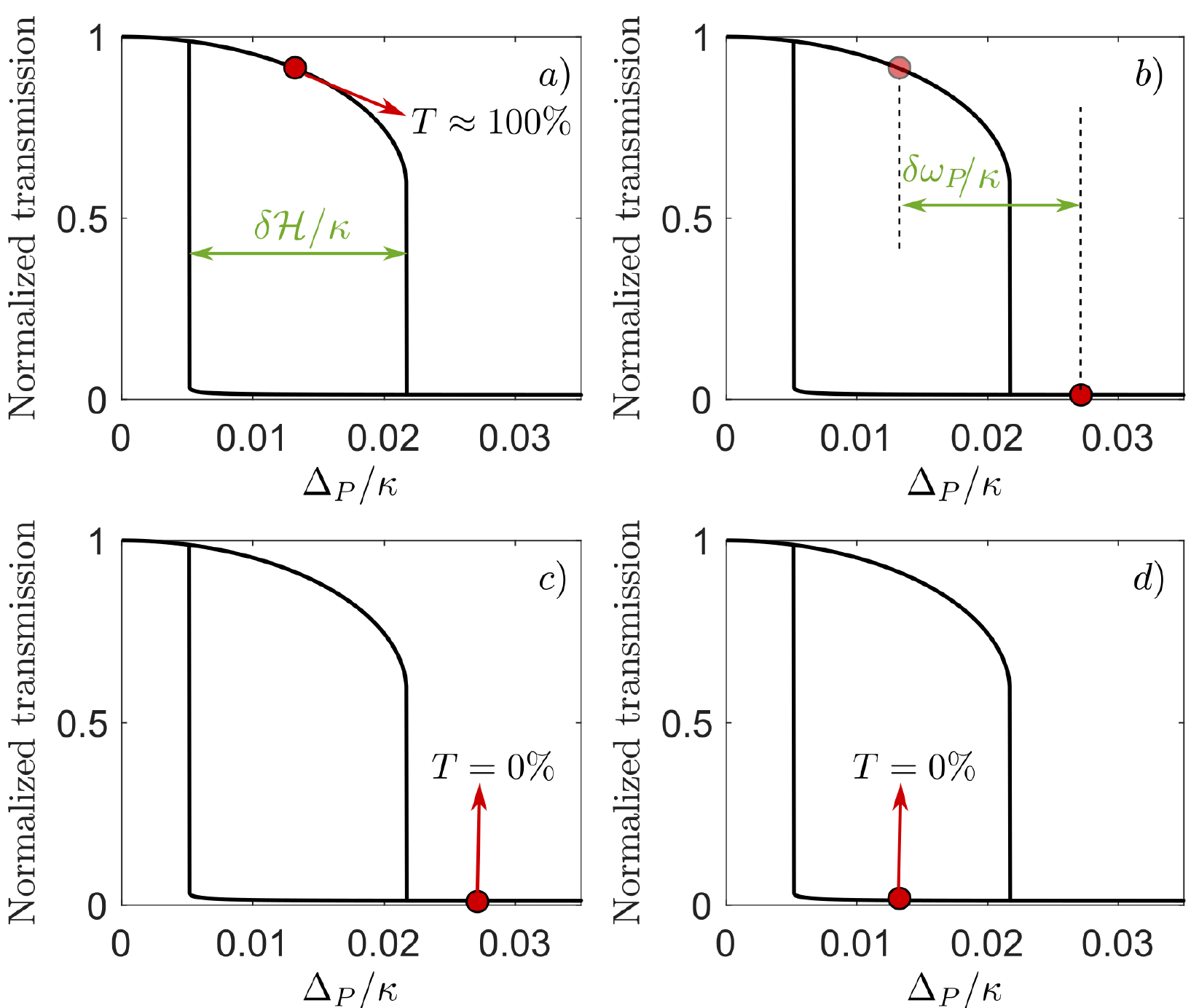}

\caption{(Color online) Schematic illustration of the operation of the proposed device. It shows the normalized transmission $\langle a^{\dagger}a\rangle/|\varepsilon/\kappa|^{2}$ as a function of the cavity-probe detuning $\Delta_{P}/\kappa$. In (a) the system is prepared in a specific state, showing a transmission $ > 90\%$, and the hysteresis has width $\mathcal{\delta H}/\kappa$. In (b), a frequency fluctuation of the probe field $\delta\omega_{P}$ is represented. When this fluctuation happens the system reaches a state where transmission is null, as represented in (c). After the fluctuation has passed through the system, it will come back to the same initial parameter configuration as in (a), but in a different state, resulting in a corresponding null transmission, as we see in (d). In this pictorial representation of the device's functioning, the hysteresis were obtained via time evolution, where only stable solutions are observed. The parameters are: $\Omega_C=0.05\kappa$, $\varepsilon=\sqrt{5}\kappa$, $N=10^3$, $C=5$, and $\Gamma_{31}=\Gamma_{32}=0.5\kappa$.}

\label{fig:6}
\end{figure}

Fig. \ref{fig:6} illustrates how the device proposed here can operate. Panel \ref{fig:6}(a) shows the normalized transmission of the system as a function of $\Delta_{P}$ and a specific state in which the system can be prepared, exhibiting a transmission $ > 90\%$. Panels (b), (c) and (d) display a sequence of states in which the system goes through due to a fluctuation $\delta\omega_{P}/\kappa$ of the probe frequency. The transmission of the system goes to zero as the $\Delta_{P}$ increases. When the probe frequency returns to its initial value, the system does not return to its initial state as showed in the \ref{fig:6}(a), thus resulting in null transmission. Therefore, the characterization of the regions where the system presents a hysteresis curve, whose width is represented by $\mathcal{\delta H}/\kappa$, is of fundamental importance in the implementation of the fluctuation detector. The same idea can be applied to detect intensity control field fluctuations.

\subsection{Frequency and intensity fluctuation detectors}

To perform the implementation of small fluctuation detectors, based on the schematic idea illustrated in Fig. \ref{fig:6}, it is necessary to prepare the system in a specific bistable regime. Here, we propose the implementation of a probe field frequency and an intensity control field fluctuations devices. Firstly let's theoretically demonstrate how to perform the measure of cavity-probe detuning fluctuation from the hysteresis profile as a function of $\Delta_{P}$. Assuming that this parameter undergoes a (for simplicity) Gaussian fluctuation during a given time interval, its time dependence is given by:
\begin{equation}
\Delta_{P}(t)=\Delta_{P}^{0}+\delta\Delta_{P}e^{-(t-t_{0})^{2}/2\sigma^{2}},
\end{equation}
where $\sigma=\sqrt{2ln(2)}FWHM$, being $FWHM$ its full width at half maximum, $\Delta_{P}^{0}$ the equilibrium cavity-probe detuning and $\delta\Delta_{P}$  the maximum detuning fluctuation, centered at $t_{0}$.

\begin{figure}
\includegraphics[width=1\columnwidth]{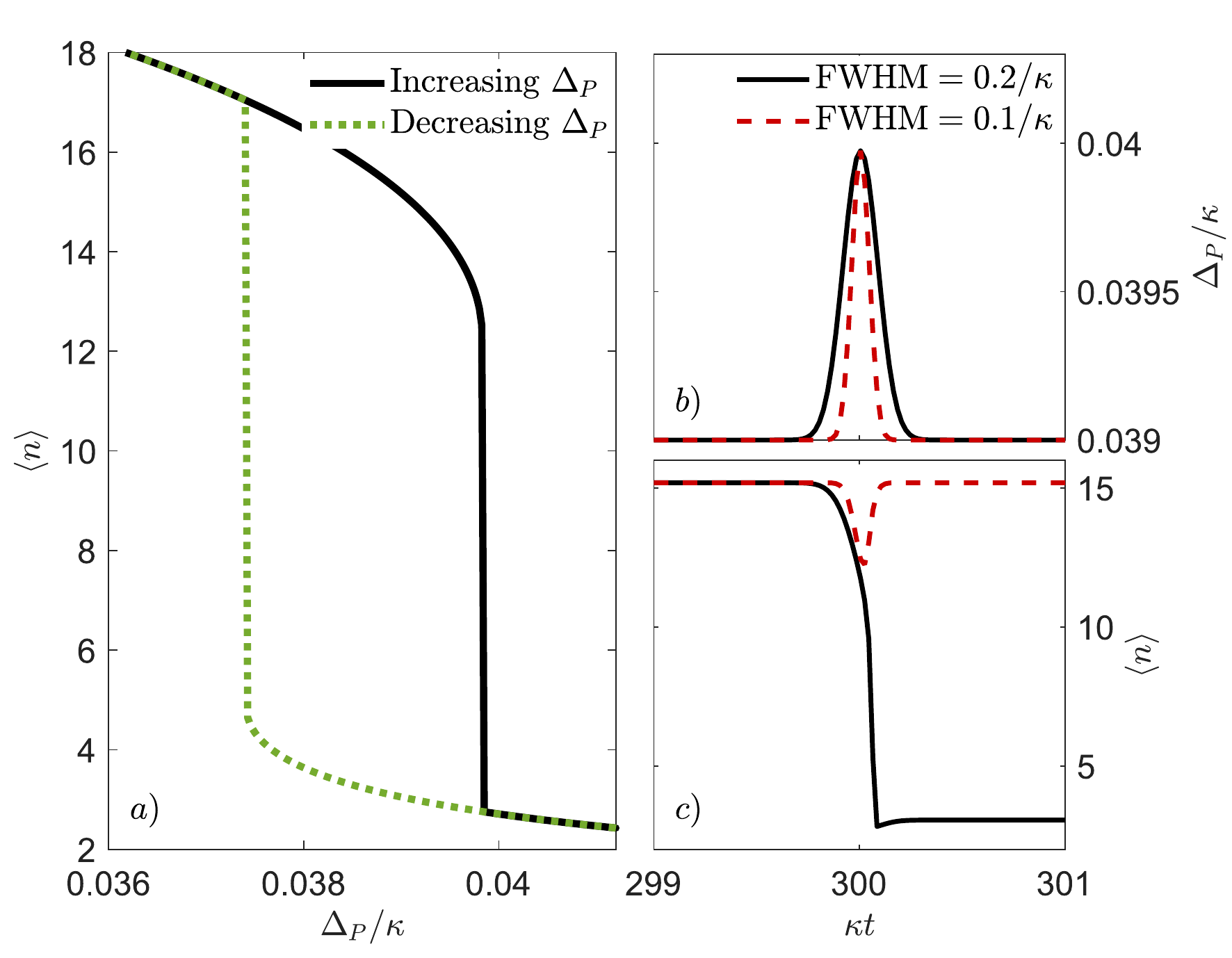}

\caption{(Color online) Schematic representation of the detector's operation, which measures small fluctuations of $\Delta_{p}$. Panel (a) shows the mean number of intracavity photons as a function of $\Delta_{P}/\kappa$ when the frequency is scanned up (solid-black line) and down (dashed-green line). Panel (b) shows the Gaussian profile of the cavity-probe detuning fluctuation in the time domain, considering two different values of full width at half maximum: $FWHM=0.1/\kappa$ (dashed-red line) and $FWHM=0.2/\kappa$ (solid-black line) and (c) shows the average number of photons $\langle n\rangle$ as a function of time ($\kappa t$) under the influence of the detuning fluctuation for the same values of $FWHM$ considered in (b). The parameters used here are: $\Gamma_{31}=\Gamma_{32}=0.5\kappa$, $\varepsilon=5\kappa$, $C=2.7$, $\Omega_{C}=0.2\kappa$, $N=10^3$, $\Delta_{p}^{0}=3.9\times10^{-2}\kappa$, $\delta\Delta_{P}=2.5\times10^{-2}\Delta_{P}^{0}$ and $t_{0}=300\kappa^{-1}$.}

\label{fig:7}
\end{figure}

In the Fig. \ref{fig:7}(a) it is shown the hysteresis curve, putting in evidence the bistable profile of the system as a function of $\Delta_{P}/\kappa$, considering the parameters $\varepsilon$ and $\Omega_{C}$ fixed. In panels \ref{fig:7}(b) and (c) the Gaussian shape of the $\Delta_{P}(t)$ fluctuation and average number of intracavity photons as a function of $\kappa t$ are plotted, respectively. For the dashed-red curves it was considered $FWHM=0.1/\kappa$ and for solid-black curves it was considered $FWHM=0.2/\kappa$. The maximum fluctuation $\delta\Delta_{P}/\kappa$ was properly adjusted as $2.5\%$ of the probe frequency detuning $\Delta_{P}/\kappa$, making clear that our protocol is able to detect a frequency fluctuation of this order.

Based on the same protocol above, our system can be also used to detect small fluctuations of the control field intensity. To this end, let's assume that the Rabi frequency of the control field has the following Gaussian time dependence:
\begin{equation}
\Omega_{C}(t)=\Omega_{C}^{0}+\delta\Omega_{C}e^{-(t-t_{0})^{2}/2\sigma^{2}},
\end{equation}
being $\Omega_{C}^{0}$ and $\delta\Omega_{C}$ the Rabi frequency value of the control field and its maximum fluctuation value, at time $t_{0}$, respectively.

\begin{figure}
\includegraphics[width=1\columnwidth]{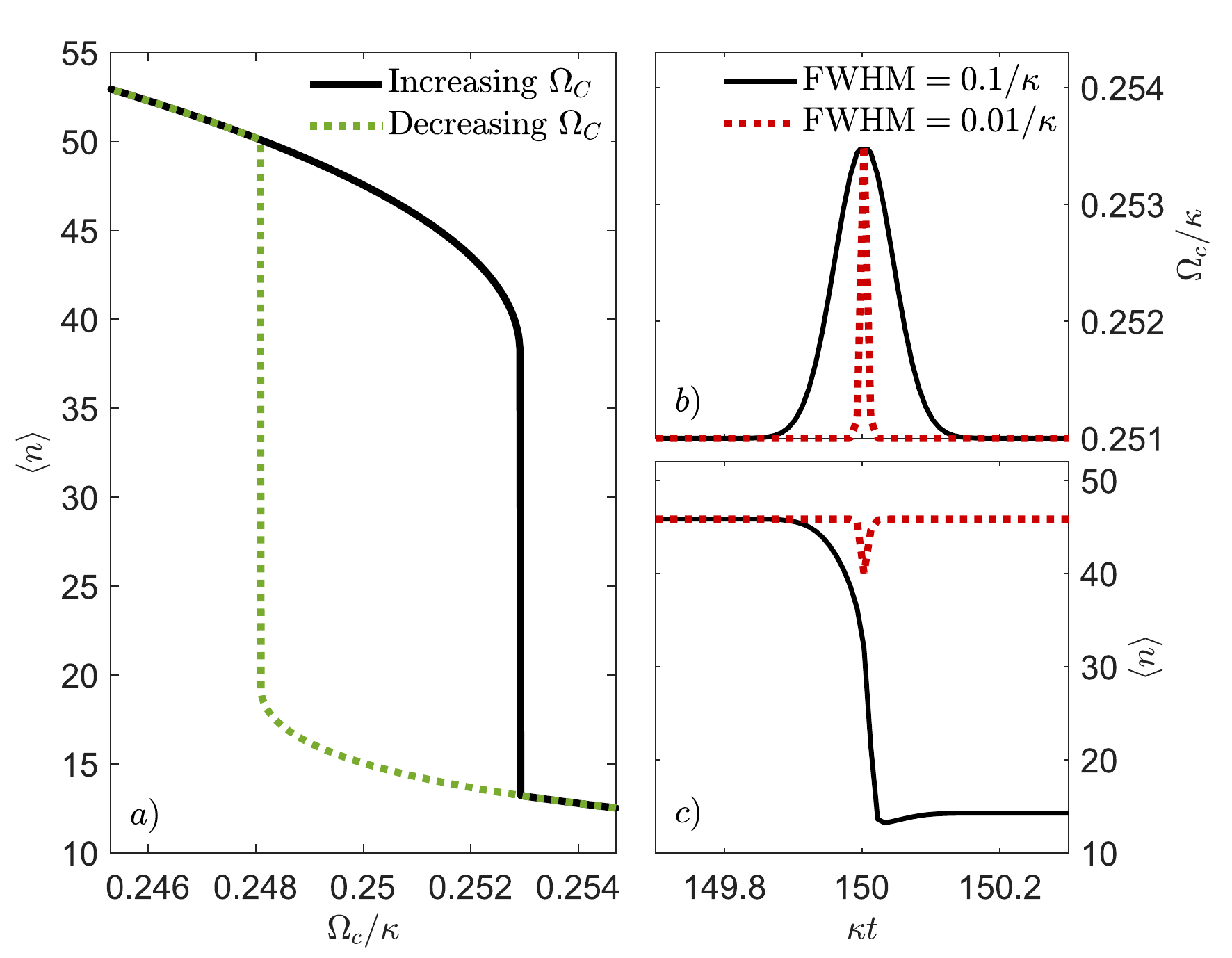}

\caption{(Color online) Schematic representation of the operation of the detector that measures small fluctuations of the Rabi frequency $\Omega_{C}$. Panel (a) shows the average number of intracavity photons as a function of $\Omega_{C}$ when the intensity of the control field is scanned up (solid-black line) and down (dashed-green line). The Gaussian shape of the Rabi frequency of the control field in the time domain is shown in panel (b), considering two different values of full width at half maximum, $FWHM=0.01/\kappa$ (dashed-red line) and $FWHM=0.1/\kappa$ (solid-black line). In (c) the average number of photons $\langle n\rangle$ is plotted as a function of time ($\kappa t$) under the influence of a $\Omega_{C}(t)$ fluctuation for the same values of $FWHM$ considered in (b). The parameters used here were: $\Gamma_{31}=\Gamma_{32}=0.5\kappa$, $\varepsilon=10\kappa$, $C=2.7$, $\Delta_{P}=0.2\kappa$, $N=10^3$, $\Omega_{C}^{0}=2.51\times10^{-1}\kappa$, $\delta\Omega_{C}=0.01\Omega_{C}^{0}$ and $t_{0}=150\kappa^{-1}$.}

\label{fig:8}
\end{figure}

In Fig. \ref{fig:8} a similar analysis of the previous result was carried out, but this time demonstrating the operation of a detector of small fluctuations in the intensity of the control field. Panel Fig. \ref{fig:8}(a) displays the hysteresis curve as a function of the Rabi frequency $\Omega_{C}/\kappa$. Panel \ref{fig:8}(b) shows the control field intensity fluctuation and Fig.~\ref{fig:8}(c) its effect on the average number of intracavity photons. Two width pulses were considered: $FWHM=0.01\kappa^{-1}$(dashed-red line) and $FWHM=0.1\kappa^{-1}$ (solid-black line). We can observe that $\langle n\rangle$ changes abruptly as the pulse fluctuation crosses the atomic sample, becoming possible to detect a fluctuation of the order of $10^{-2}\Omega^0_{C}$. 

It is worth mentioning that by using this procedure it is possible to further increase the detectors precision, since through suitable adjustments in the system parameters and, consequently, a proper control over the bistable profile of the system, one could obtain a greater transmission contrast in response to the fluctuation of the parameter they have interest in measuring. One possible limitation of such detectors is not related to the parameter in which the fluctuation is observed or even to the measurement precision, but to how fast does the fluctuation occurs. In Figs. \ref{fig:7} and \ref{fig:8}, we show an example of how two different pulse's FWHM can affect the device's functionality. If the fluctuation is fast enough, the system has no time to respond, thus no transmission switching is observed. This, however, will depend on a series of parameters, such as the number of photons, the number of atoms, the coupling strengths and the decay rates, which are specific to the desired device application.

\textemdash{}

\section{Conclusion}
We have theoretically investigated the optical bistability phenomenon in a three-level atomic system in the $\Lambda$-type configuration. One of the atomic transition is coupled to a cavity mode while the other is induced by a classical control field. The optical bistability was already investigated experimentally in this system, which in turn present a great control of the hysteresis in the transmission spectrum of the cavity  \citep{Wang2002,Brown2003,Joshi2003a,Joshi2010}. But here we have shown that this system allows for other kinds of bistability. Instead of scanning the intensity of the pumping field, we have shown that the scanning of either the Rabi frequency of the control field or the frequency of the pumping field also presents some bistable behavior, even for a very small average number of intracavity photons. The hysteresis in our system can be highly controllable through external parameters as the frequencies and intensities of the laser fields, and its width can be very narrow, thus been very useful for applications in optical devices as detectors of small fluctuations of the intensity or frequency of classical fields, as shown here. Finally, we have shown numerically that our system predicts bistable behavior for Cooperativity $C<4$ (in our simulations we could observe bistability even for $C=1.75$) which is below the threshold for two-level systems. An analytical solution for our system of equations could reveal what is indeed the smallest possible value for the Cooperativity which allows for a bistable behavior. However, our result is already of fundamental importance since it shows that $C=4$ is not a fundamental limit. 

\begin{acknowledgments}
C.J.V.-B., H. S. Borges and J.A.S. acknowledge support from the São Paulo Research Foundation (FAPESP), grants 2013/04162-5, 2014/12740-1, 2019/11999-5 and 2015/21229-1. C.J.V.-B and M.H.O. thank the support from the National Council for Scientific and Technological Development (CNPq) grants 307077/2018-7 and 141247/2018-5. The authors also thanks the support from the Brazilian National Institute of Science and Technology for Quantum Information (INCT-IQ/CNPq) Grant No. 465469/2014-0 and by the Coordenação de Aperfeiçoamento de Pessoal de Nível Superior - Brasil (CAPES) - Finance Code 001. 
\end{acknowledgments}

\bibliographystyle{unsrt}
\bibliography{ref}

\end{document}